\documentclass[journal]{IEEEtran}
\usepackage{xcolor,soul,framed} 

\colorlet{shadecolor}{yellow}
\usepackage[pdftex]{graphicx}
\graphicspath{{../pdf/}{../jpeg/}}
\DeclareGraphicsExtensions{.pdf,.jpeg,.png}

\usepackage[cmex10]{amsmath}
\usepackage{array}
\usepackage{commath}
\usepackage{mdwmath}
\usepackage{mdwtab}
\usepackage{eqparbox}
\usepackage{url}
\usepackage{amsmath}
\usepackage{bbm}
\usepackage{yfonts}
\usepackage{dsfont}
\usepackage{ amssymb }
\usepackage[T1]{fontenc}
\usepackage[utf8]{inputenc}
\usepackage{authblk}
\usepackage{amsthm,amssymb}

\usepackage[utf8]{inputenc}
\newcommand{\argminF}{\mathop{\mathrm{argmin}}\limits} 

\newcommand{\argmaxF}{\mathop{\mathrm{argmax}}\limits} 
\title{Channel Estimation for Diffusive MIMO Molecular Communications}
\author{S. Mohammadreza Rouzegar, Umberto Spagnolini}
\affil{Dipartimento di Elettronica, Informazione e Bioingegneria, Politecnico di Milano, Milan, Italy}
\affil{Email: seyedmohammadreza.rouzegar@mail.polimi.it, umberto.spagnolini@polimi.it}

\begin{document}
\maketitle 
\newcommand{\vect}[1]{\boldsymbol{#1}}
 \begin{abstract}
 In diffusion-based communication, as for molecular systems, the achievable data rate is very low due to the slow nature of diffusion and the existence of severe inter-symbol-interference (ISI). Multiple-input multiple-output (MIMO) technique can be used to improve the data rate. Knowledge of channel impulse response (CIR) is essential for equalization and detection in MIMO systems. This paper presents a training-based CIR estimation for diffusive MIMO (D-MIMO) channels. Maximum likelihood and least-squares estimators are derived, and the training sequences are designed to minimize the corresponding Cramér-Rao bound. Sub-optimal estimators are compared to Cramér-Rao  bound to validate their performance. 
 \end{abstract} 
 \begin{IEEEkeywords}
 Molecular communication, Diffusive Multiple-input Multiple-output, Channel impulse response, Cramér-Rao  Bound, Training Sequence Design
 \end{IEEEkeywords}  

\section{Introduction}
Molecular communication (MC) is a bio-inspired solution of communication at nano-scale \cite{nakano2005molecular,farsad2016comprehensive}. Conventional communication systems transfer information using electromagnetic waves. At nano-scale, antennas suffer the constraint of being at comparable scale of electromagnetic wavelength. Additionally, using electromagnetic wave for nanomachines can be detrimental in some environments, such as inside a body where electromagnetic radiation can be harmful for health. Hence, MC can be a preferred solution for communication among nanomachines to build a nanonetwork, so they perform complex tasks which could not be possible individually \cite{akyildiz2015internet,nakano2014molecular}.

In MC, bio-nanomachines communicate through exchanging molecules. In fact, the simplest system needs a transmitter to send the information molecules, and a receiver to collect them. Information molecules diffuse toward the receiver using Brownian motion resulting from their collision with the molecules in the fluid \cite{kadloor2012molecular,nakano2013molecular}. Information can be encoded to the different properties of molecules, such as their concentration \cite{kuran2011modulation}, number \cite{noel2014optimal}, type \cite{cobo2010bacteria}, and time of release \cite{farsad2016capacity}. 

One of the main challenges of MC is to deal with the long tail of diffusive propagation that causes severe inter-symbol-interference (ISI). One can increase the symbol interval time to eliminate the ISI, but the need for higher data rate justifies the optimization of the symbol interval time to have few channel taps due to the ISI \cite{kim2014symbol}. Even if one optimizes the symbol interval time, the  slow nature of diffusion makes the data rate still low. Using multiple-input multiple-output (MIMO) technique is a widely investigated solution to address this problem \cite{koo2016molecular,meng2012mimo} and can be adopted for MC. 

In this paper, we assume that information is encoded in the number of molecules observed at the receiver. Therefore, we define the channel impulse response (CIR) as the expected number of molecules observed at the receiver domain at time $t$ denoted as $\bar{\vect{c}}(t)$ after instantaneous release of molecules at $t=0$ \cite{jamali2016channel}. We assume the number of molecules at the receiver follows the Poisson distribution as introduced in \cite{jamali2016channel,arjmandi2013diffusion} and the CIR is their mean number of molecules.

The goal of this paper is to gain insight to the estimation of the CIR of   D-MIMO communication channel. In \cite{meng2012mimo}, authors investigated various diversity technique in D-MIMO communication assuming full knowledge of CIR.  The authors of \cite{koo2016molecular}, modeled the $2\times 2$ D-MIMO channel by fitting a curve to the simulated data. Goal of this paper is to define a training-based channel estimation for D-MIMO  molecular communication.
In \cite{jamali2016channel}, authors introduced the channel estimation for the  Poisson  MC channel and proposed the estimators of CIR in single transmitter and single receiver system. Novelty of this work can be considered the introduction of a system model and notation (Section 2) that allow us to estimate the CIR of $M\times M$ D-MIMO Poisson channel. We extended the steps of R. Schober et. al. \cite{jamali2016channel} to D-MIMO channel estimation (Section 3) by accounting for the inter-link diffusive interference. Furthermore, we propose in Section 4, a D-MIMO specific method for designing the training sequence that minimize the Cramér-Rao bound (CRB) at all receivers simultaneously.

\section{System model } 
We consider a $M\times M$ D-MIMO system for MC as shown in Fig. 1. The system consists of $M$ pair of transmitters denoted as $Tx_i $, and receivers $Rx_j$, where $i,j \in \{ 1,2,3...M\} $.
We assume that  all transmitters emit the same type of molecules. Each transmitter emits a known number of molecules $N$ at the beginning of each symbol intervals. The molecules diffuse in the environment and some of them reach the $M$ receivers.
Transmitters and receivers are not fixed in their position and could slightly move on fluid where molecules diffuse, so the CIR changes over time. We assume a block-type communication and we estimate the CIR at the beginning of every block by sending a properly designed training sequence, and we assume that the D-MIMO channel does not vary during each block. Then, the estimated CIR is used for equalization at the receiver over the rest of the block.\par 
Transmitters modulate the molecules density using concentration shift keying (CSK) and the receivers count the number of molecules at the time of sampling. As customary, we set the sampling time so that the number of molecules at receivers  for the corresponding transmitters is maximized. 
 As shown in Fig. 2, the channel has memory and due to the inter-symbol-interference (ISI), the receiver counts the molecules from previous samples of the corresponding transmitter. Similarly,  the molecules from the current and previous samples of the non-corresponding transmitters are known as inter-link-interference (ILI). Number of channel taps (L) for each link is related to the system geometry, configuration and sample interval. Specifically, we can eliminate the ISI and ILI by making the sample interval large enough and putting each pair of transceivers far enough from the other pairs. However, this case is not considered due to the demands for high data rate per unit of space. Hence, we have to face the ISI and ILI in the MC system and try to mitigate their effects. The observed number of molecules at sampling time $k$ and receiver $j$ is
\begin{figure} 
  
  \includegraphics[width=3.4in]{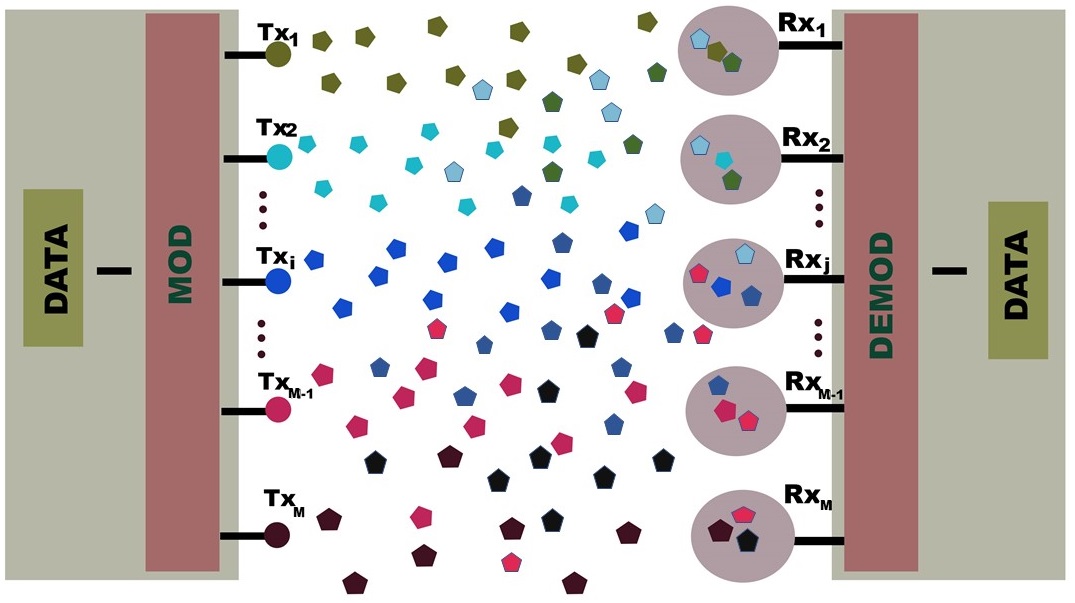}\\
  \caption{ Topological model for $M\times M$ D-MIMO system. The M transmitters $(Tx_1,...,Tx_M)$ release the same type of molecules (pentagon shape), and molecules here have different colors according to the corresponding transmitter.    }\label{N_N}
 
\end{figure}
\begin {equation}\label{first equation}
y_j[k]=\sum_{i=1}^{M}\sum_{\ell=0}^{L-1} c_{ij}[\ell,k]
x_i[k-\ell]  +v_j[k]
\end{equation}
where $L$ is the  memory taps, $c_{ij}[\ell,k]$ is a random variable and denotes to the number of molecules observed at the receiver $j$ from transmitter $i$ due to the release of $N$ molecules at the time interval $[k-\ell]$. Case $i=j$ refers to the paired transmitter-receiver, otherwise it refers to the inter-link interference. $x_i[k] \in \{0,1\}$ is the transmitted symbol at the time interval $k$ from transmitter $i$. The number of molecules  $c_{ij}[\ell ,k]$ can be approximated as a Poisson random variable with a mean value $\bar{c}_{ij}[\ell]$: $c_{ij} [\ell,k] \sim Poiss \, (\bar{c}_{ij} [\ell]) $. Additionally,
$v_j[k]$ is the number of external noise molecules detected at the receiver $j$ at time interval $k$. Noise molecules could originate from the remaining channel taps from all transmitters not considered in  model, and any external source. Hence, we can consider the noise as a Poisson with a mean $\bar{v}_j$: $v_{j}[k] \sim Poiss \, (\bar{v}_{j})$  \cite{noel2014unifying}. \par

\begin{figure}
  
  \includegraphics[width=3.6in]{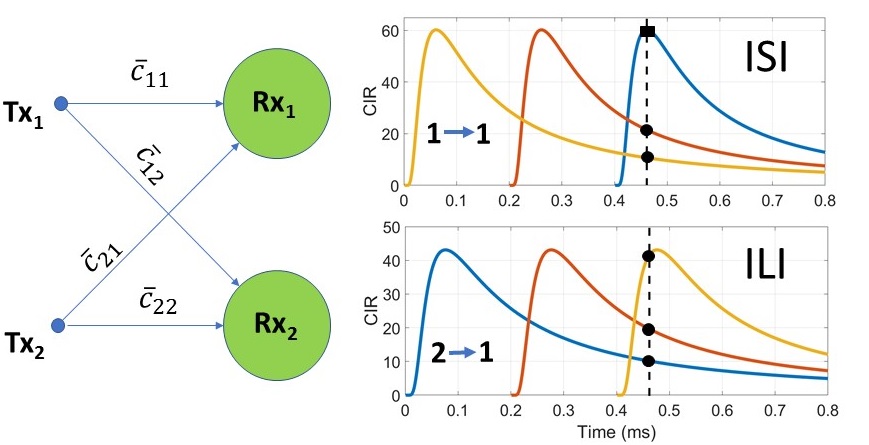}\\
  \caption{ Impulse response, $\bar{c}_{ij}(t)$, of a $2\times 2$ D-MIMO system at $Rx_1$ vs time, for 3 emissions of molecules with time spacing $0.2 \,ms$: ILI and ISI are black dots. }\label{MIMO_channel}
 
\end{figure}

\section{Problem Definition}
Assume that $\vect{x}_i= [x_i[1],x_i[2],..., x_i[K]]^T$ is a binary training sequence with length $K$ for transmitter $i$. To avoid edge effect due to the ISI, in CIR estimation we employ $y_j[k]$ for $ k\geq L$. Therefore, the $K-L+1$ samples are used for CIR estimation of the $j$-th receiver. The mean number of molecules at receiver $j$ at time $k$ is
\begin {equation}\label{mean1 equation}
\bar{y}_j[k]= \mathds{E} \, \, \{ y_j[k] \}=
\sum_{i=1}^{M}\sum_{\ell=0}^{L-1} \bar{c}_{ij}[\ell] 
x_i[k-\ell]  + \bar{v}_{j}
\end{equation}
which $L\leq k\leq K$. Eqn. (\ref{mean1 equation}) can be written compactly as:
\begin{equation}\label{y j receiver}
\bar{y}_j[k]=\vect{X}^T[k]\vect{\bar{C}}_j
\end{equation}
where the following notations are used:
$$\vect{x}[k]=[x_1[k],x_2[k],...,x_M[k]]^T$$
$$\vect{X}[k]=[\vect{x}^T[k],\vect{x}^T[k-1],...., \vect{x}^T[k-L+1],1]^T$$
$$\vect{\bar{c}}_j[\ell]= [\bar{c}_{1j}[\ell],\bar{c}_{2j}[\ell], ...,\bar{c}_{Mj}[\ell]]^T$$
$$\vect{\bar{C}}_j=[\vect{\bar{c}}_j^T[0],\vect{\bar{c}}_j^T[1],...,\vect{\bar{c}}_j^T[L-1],\bar{v}_j]^T$$
here, $\vect{x}[k]$ is a $M\times 1$ vector denotes to the training sequence of all transmitters at time $k$,  $\vect{\bar{c}}_j[\ell]$ is a  vector with dimension $M\times 1$ denoting the expected number molecules at receiver $j$ at time $k$ due to the transmission of $N$ molecules at time $k-\ell $ from  transmitter $Tx_i$, and $\vect{x}[k-\ell]$ is its corresponding training sequence. Additionally, $\vect{\bar{C}}_j$ is a $(ML+1)\times 1$ vector gathering all channel memory taps  of receiver $j$ due to the ISI and ILI from all transmitters and also noise $\bar{v}_j$, and $\vect{X}[k]$ with dimension $(ML+1)\times 1$ is its corresponding training sequence vector at time interval $k$. \par 
The expected number of molecules at receiver $j$ during the time intervals $L \leq k\leq K$ is defined as $\vect{\bar{y}}_j= [\bar{y}_j[L],\bar{y}_j[L+1]...,\bar{y}_j[K]]^T $, and the D-MIMO relation for receiver $j$ is written
\begin{equation} \label{y_j vector}
\underset{ (K-L+1)\times 1}{\vect{\bar{y}}_j}
=\underset{(K-L+1)\times (ML+1)}{\vect{X}^T} \, \,\,\underset{(ML+1)\times 1}{\vect{\bar{C}}_j}
\end{equation}
where $\vect{X}$ is a $(ML+1)\times (K-L+1)$ convolution matrix of training sequences including memory of previous samples due to the channel taps, and it is defined as
\begin{equation}
\vect{X}=[\vect{X}[L],\vect{X}[L+1],...,\vect{X}[K]]
\end{equation}

Finally, we define $\vect{\bar{Y}}=[\vect{\bar{y}}_1,\vect{\bar{y}}_2...,\vect{\bar{y}}_M]$ and we compactly write the global D-MIMO relation into
\begin{equation}\label{MIMO eqn}
\underset{(K-L+1)\times (M)}{\vect{\bar{Y}}}=\underset{(K-L+1)\times (ML+1)}{\vect{X}^T} \,\,\, \underset{(ML+1)\times M}{\vect{\bar{C}}}
\end{equation}
where  $\vect{\bar{C}}$ is the $(ML+1)\times M$ global channel matrix, and it is defined as 
\begin{equation}
\vect{\bar{C}}=[\vect{\bar{C}}_1,\vect{\bar{C}}_2,...,\vect{\bar{C}}_M].
\end{equation}

The matrix of all the observed number of molecules at the $M$ receivers contain the Poisson random variables with mean equal to $\vect{\bar{Y}}$:
\begin{equation}\label{Poiss Y}
\vect{Y}=Poiss \, (\vect{\bar{Y}})
\end{equation}
which means each entry of the observed matrix $\vect{Y}$, is Poisson random variable with mean equal to the corresponding entry of  $\vect{\bar{Y}}$. \par 
The probability density function (PDF) of all observations at all receivers are the product of the Poisson distribution of each observation at each receiver 
\begin{equation}\label{PDF}
f_{\vect{Y}} \, (\vect{Y}|\vect{\bar{C}},\vect{X}) = \prod_{k=L}^K \prod_{j=1}^{M}
\frac{\big(\vect{X}^T[k]\vect{\bar{C}}_j)^ {y_{j}[k]} \, \exp (- \vect{X}^T[k]\vect{\bar{C}}_j)}
{y_{j}[k] \,!}
\end{equation}

According to (\ref{y_j vector}), we can analyze the performance of each receiver independently to make sure that all $M$ receivers are simultaneously working optimally. Therefore, the PDF of the observations of $j$-th receiver is
\begin{equation}\label{PDF j}
f_{\vect{y}_j} \, (\vect{y}_j|\vect{\bar{C}}_j,\vect{X}) = \prod_{k=L}^K 
\frac{\big(\vect{X}^T[k]\vect{\bar{C}}_j)^ {y_{j}[k]} \, \exp (- \vect{X}^T[k]\vect{\bar{C}}_j)}
{y_{j}[k] \,!},
\end{equation}
for maximum likelihood estimation below.

\section{D-MIMO Channel Estimation}
In this section sub-optimal maximum likelihood (ML) and least squares (LS) estimators for the D-MIMO system are derived. Then,  CRB for each receiver is computed.

\subsection{Maximum Likelihood estimator}
Maximum likelihood (ML) D-MIMO CIR estimator finds the CIR which maximize the likelihood of observation vector $\vect{y}_j$ 
\begin{equation}\label{ML}
\begin{split}
\hat{\bar{\vect{C}}}_j^{\mathrm{ML} } = \argmaxF_{{\bar{\vect{C}}_j \geq 0}}  \,  f_{\vect{y}_j} \, (\vect{y}_j|\vect{\bar{C}}_j,\vect{X})\\ =\argmaxF_{{\bar{\vect{C}_j} \geq 0}}  \,  \mathcal{L} _{\vect{y}_j} \, (\vect{y}_j|\vect{\bar{C}}_j,\vect{X}) 
\end{split}
\end{equation}
where the log likelihood function and is given by
\begin{equation} \label{log}
\mathcal{L} _{\vect{y}_j} \, (\vect{y}_j|\vect{\bar{C}}_j,\vect{X}) =\sum_{k=L}^{K} \Big [  - \vect{X}^T[k] \vect{\bar{C}}_j + y_{j}[k] \ln (\vect{X}^T[k] \vect{\bar{C}}_j)\,
\Big ]
\end{equation}

The ML estimate of the CIR for the D-MIMO channel at receiver $j$ is obtained by solving a system of non-linear equations given below \cite{jamali2016channel}:
\begin{equation} \label{ML equations}
\sum_{k=L}^{K} \Big [ \frac{y_{j}[k]\vect{X}[k]}{\vect{X}^T[k]\vect{\bar{C}}_j} -\vect{X}[k] \,
\Big ] = \vect{0}
\end{equation}
$Remark 1$: We note entries of $\vect{\bar{C}}_j$ are positive semidefinite. However, ML estimator could estimate a negative value for some elements of the $\vect{\bar{C}}_j $. Sub-optimal solution is to set to zero all the negative entries of the estimated CIR.  This heuristic approach was adopted for single link MC \cite{jamali2016channel}, and showed therein a negligible loss of performances compared to the optimal ML. Therefore, sub-optimal solution of (\ref{ML equations}) is highly preferred in D-MIMO channels due to its simplicity. 
\subsection{Least Squares Estimator}

The least squares (LS) method chooses $\vect{\bar{C}}$ which minimizes the sum of the square errors at all receiver from the observation vector $\vect{Y}$, 
\begin{equation}\label{arg LS}
\hat{\bar{\vect{C}}}^{\tiny LS} = \argminF_{{\bar{\vect{C}} \geq 0}} \, \| \vect{\epsilon} \| ^2.
\end{equation}
where $\vect{\epsilon} = \vect{Y} - \mathbb{E} \{ \vect{Y}\} =
\vect{Y}- \vect{X}^T \vect{\bar{C}} $. 
The LS estimate of the CIR for D-MIMO channel is
\begin{equation} \label{LS Estimator}
\hat{\bar{\vect{C}}}^{\tiny LS} = \Big[  (\vect{X} \, \vect{X}^T) ^{-1}   \, \vect{X} \, \vect{Y}    \Big].
\end{equation}
Minimization of (\ref{arg LS}) is a constrained optimization problem with constraint $\vect{C} \geq 0$ for entries. In case there exist a stationary point, this is the global optimum solution. In case the stationary point does not exist, sub-optimal solution is to set all negative elements of $\vect{C}$ to zero. Optimal solution for (\ref{arg LS}) is introduced in \cite{jamali2016channel}, and the authors showed that for K large, there exist a stationary point, and for small lengths, the performance loss is very negligible. Again, we prefer the sub-optimal solution for D-MIMO system due to its simplicity. 
\subsection{Cramér-Rao Bound}
The Cramér-Rao bound (CRB) sets the lower bound on the covariance of any unbiased estimator of a deterministic parameters. Let $\hat{\bar{\vect{C}}}_j$ be the unbiased estimator of $\bar{\vect{C}}_j$, the CRB sets the bound of the covariance
\begin{equation}\label{CRB inequality}
cov (\hat{\bar{\vect{C}}}_j) \succeq  \vect{I}^{-1} (\bar{\vect{C}}_j)
\end{equation}
where $ \vect{I} (\bar{\vect{C}}_j)$ is the Fisher information matrix of  $\bar{\vect{C}}_j$ and is given by
\begin{equation} \label{Fisher}
\vect{I} (\bar{\vect{C}}_j)= \mathbbm{E} _{\vect{y}_j} \{ 
- \nabla _{\vect{\bar{C}}_j\vect{\bar{C}}_j}^2 \mathcal{L} _{\vect{y}_j} \, (\vect{y}_j|\vect{\bar{C}}_j,\vect{X})\} 
\end{equation}
Therefore, the CRB at receiver $j$ is given by
\begin{equation}\label{CRB}
CRB_j=tr\{\vect{I}^{-1} (\bar{\vect{C}}_j)\}=
tr\Bigg\{ \Bigg[ \sum_{k=L}^{K} \frac{\vect{X}[k]\vect{X}^T[k]} {\vect{X}^T[k] \vect{\bar{C}}_j } \Bigg] ^{-1}
\Bigg\}.
\end{equation}
Notice that (\ref{CRB inequality}) implies that the difference $cov (\hat{\bar{\vect{{C}}}}_j)-\vect{I}^{-1} (\bar{\vect{C}}_j)\succeq 0$ is positive semidefinite. 
\section{Training Sequence Design} 
In this section, we present a method for designing the training sequences for estimating the CIR of a D-MIMO channel. As shown in (\ref{CRB}), the CRB for a given system is a function of training sequences. Therefore, we can find a set of training sequences that minimize the CRB of all receivers. In other words, the CRB of a specific receiver depends on the training sequence of all transmitters which have interference with it. In general, for a $M\times M$ D-MIMO system, we have to design $M$ different training sequences to minimize the CRB of all receivers simultaneously. However, in practice we do not need to design $M$ training sequences, because ILI for far transmitters is negligible and thus  we neglect their interference channels but consider them as an augmented noise source in $v_j$.
In order to find a suitable set of training sequences that simultaneously minimize all CRBs, we consider following constraints: 1) the training sequences should be molecularly efficient by minimizing the fraction of molecules used for channel estimation, and 2) transmitters can not be silent for many consequent intervals. In detail, for a training sequence of length K, we consider sequences with maximum $K/2$ ones, consequently transmitting maximum $NK/2$ molecules, and the maximum consequent zeros are considered 4 time intervals. $\mathcal{X}$ is the sets of all possible training sequences that meet the above criteria. 
\begin{equation}\label{sequence}
[\vect{x}_1,\vect{x}_2,...,\vect{x}_M]=
\argminF_{\vect{x}_i\in \mathcal{X}} \{ CRB_1, \dots CRB_M\}
\end{equation}
\begin{figure}
  
  \includegraphics[width=3.55in]{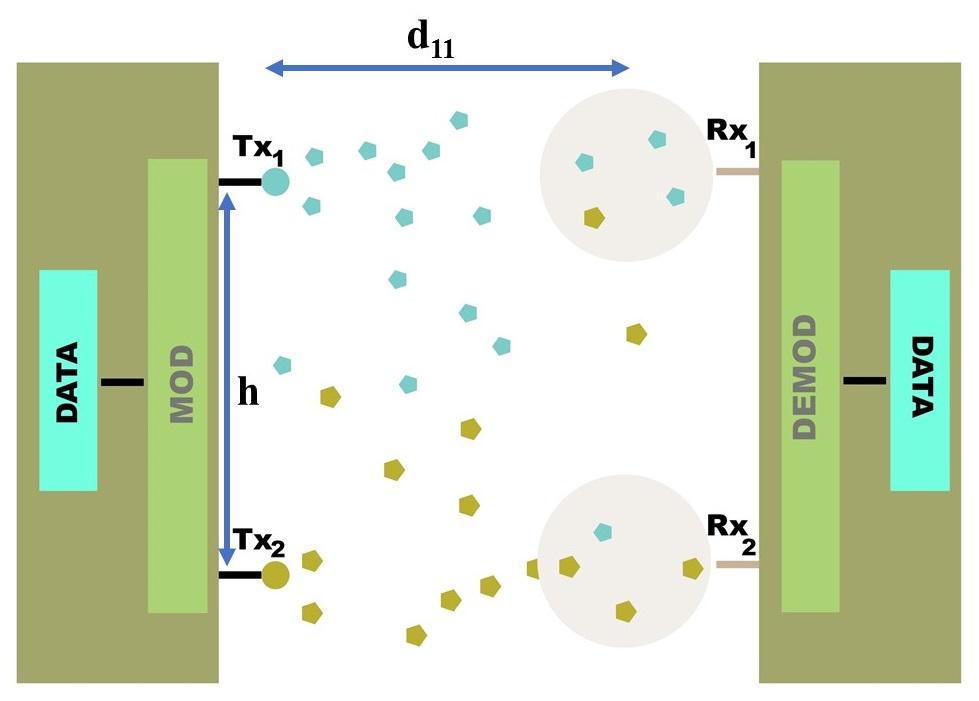}\\
  \caption{Topological model for a $2\times 2$ D-MIMO system. Both transmitters use the same information molecules (pentagon shapes), and molecules here have different colors according to the corresponding transmitter.}\label{2_2MIMO}
 
\end{figure}
\textit{Remark 2:} Accuracy of CIR estimation depends on the training sequence length, hence $K$ should chosen carefully. For large $K$, it is difficult to search among all suitable sets and find the optimum ones. Therefore, we look for an optimum training sequence with smaller length $K_1$, and we concatenate it to build  a longer training sequence of length $K$. We observed that if $K_1$ is  wisely selected, concatenating would not impair the performances.

\section{Performance Evaluation of the Designed D-MIMO System}
In this section, we present a $2\times 2$ D-MIMO configuration and we compare the performances of the estimators introduced in this paper for ON-OFF keying signaling. We have generated the CIR according to the analytical models proposed in \cite{noel2014optimal,kadloor2012molecular} for MC systems. However, there is no constraint in the value of CIR to be estimated. Diffusion coefficient value is $10^{-9} m^2/s$ and it is compatible to the normal values of diffusion of most of molecules in water at room temperature.  Choosing bit interval time is a trade-off between bit rate and total number $LM$ of ISI and ILI of the channel memory to be estimated. Bit interval time is $T_{int}=0.2 \, ms$, all transmitters release $N=10^5$ molecules and   the receivers counts once the number of molecules  per each symbol  at time which CIR of the pair transmitter is expected to be maximum. For simplicity, the mean of noise  is chosen as $\bar{v_j}= 0.3 \, \bar{ {c}}_{jj} (0)$. The number of channel taps for both ISI and ILI link are considered $L=3$, so $\bar{c}_{ij}[L] \leq 0.05 \, \bar{c}_{jj}[0] $.
\begin{figure}
  
  \includegraphics[width=3.55in]{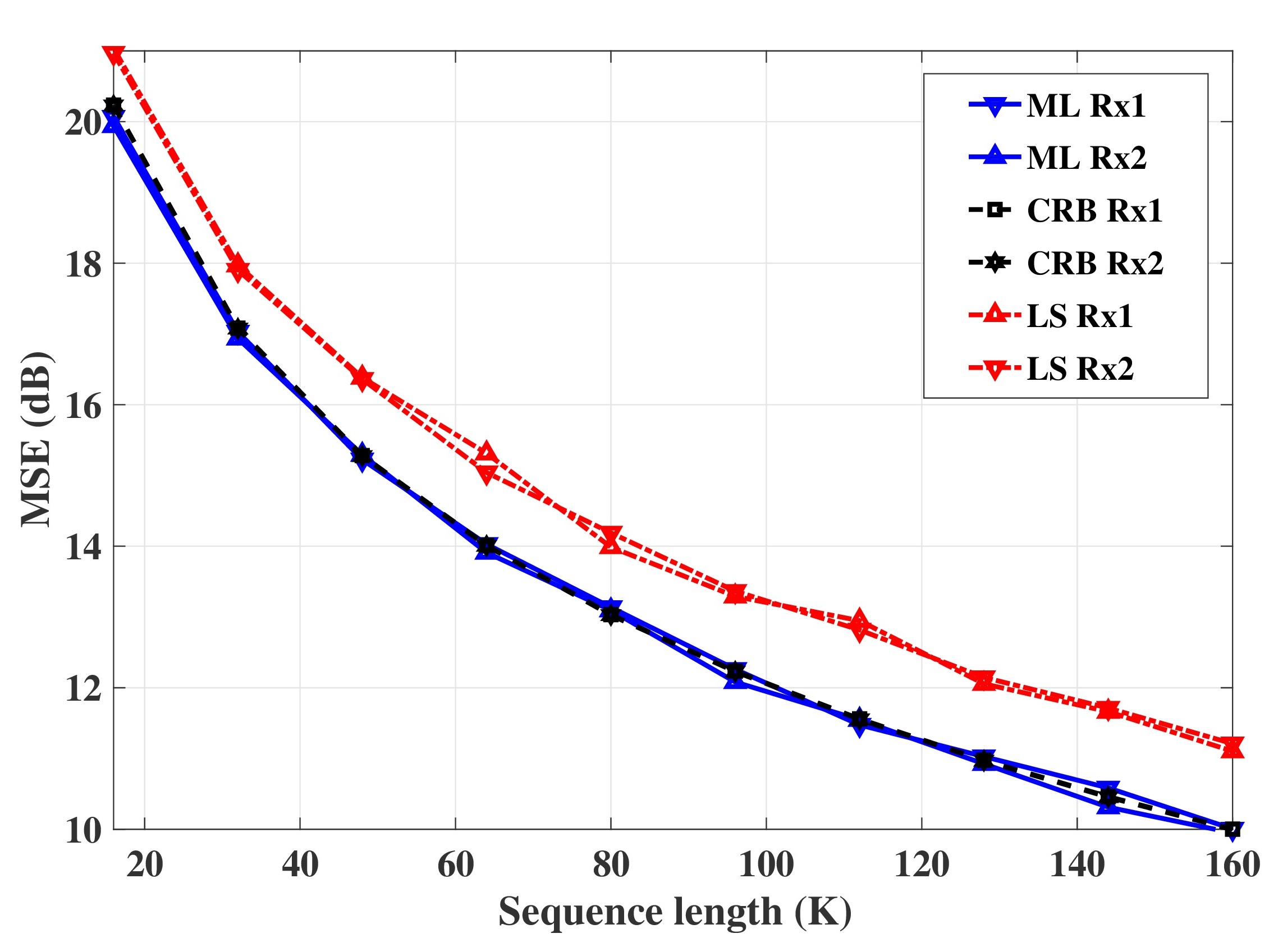}\\
  \caption{Comparison of ML and LS estimators to CRB in terms of MSE in dB vs. the training sequence length K with $L=3$.}\label{ML_LS}
 
\end{figure}

The $2\times 2 $ D-MIMO system is shown in Fig. \ref{2_2MIMO}. We have assumed that the distance between transmitter and receiver is $d= 400 \, nm$ and the inter-distance is $h=200 \, nm$. Spherical receiver with radius $50\, nm$ is assumed. Positions of the mentioned entities are fluctuating: $P_{Tx_1}=(0+\delta x_1 , 0+\delta y_1, 0+\delta z_1) , P_{Tx_2}=(0+\delta x_2 , h+\delta y_2, 0+\delta z_2) , P_{Rx_1}=(d+\delta x_3 , 0+\delta y_3, 0+\delta z_3) , P_{Rx_2}=(d+\delta x_4 , h+\delta y_4, 0+\delta z_4) $, with $\delta_{x,y,z} \sim \mathcal{N}(0,\sigma^2)$, and $\sigma^2 = 50\, nm$. Since the $Tx$ and $Rx$ are not fixed in the position, the channel is varying in time. While the entities are fixed, the CIR at the receivers are:  $\vect{\bar{C}}_{1}=[60.21, 41.58  ,   9.11 ,   8.71 ,  3.83 ,  3.74, 18.06 ]^T$ and $\vect{\bar{C}}_{2}=[ 41.58  ,60.21,8.71 ,   9.11 ,  3.74,   3.83 ,   18.06]^T.$ We note that in each realization the CIR is different, $\vect{\bar{C}}_{1} \neq \vect{\bar{C}}_{2}$, because the position of transmitters and receivers are changing with normal distribution with $\sigma^2 = 50\, nm$.
    The training sequences are designed  according to (\ref{sequence}) with the length $K_1=16$ and they are $\vect{x}_1=[1,1,1,0,0,0,0,1,0,1,0,1,1,0,0,1]^T$ and $\vect{x}_2=[1,1,1,0,1,0,0,0,1,1,1,0,0,0,0,1]^T$. Longer training sequences are constructed by concatenating these training sequences as detailed in section 5. \par 
    In this problem, each receiver has to estimate $LM+1=7$ variables, for a total of 14 variables. The results in Fig. \ref{ML_LS}, are Monte Carlo simulations with 1000 random CIR realizations.  \par
\begin{figure}
  
  \includegraphics[width=3.55in]{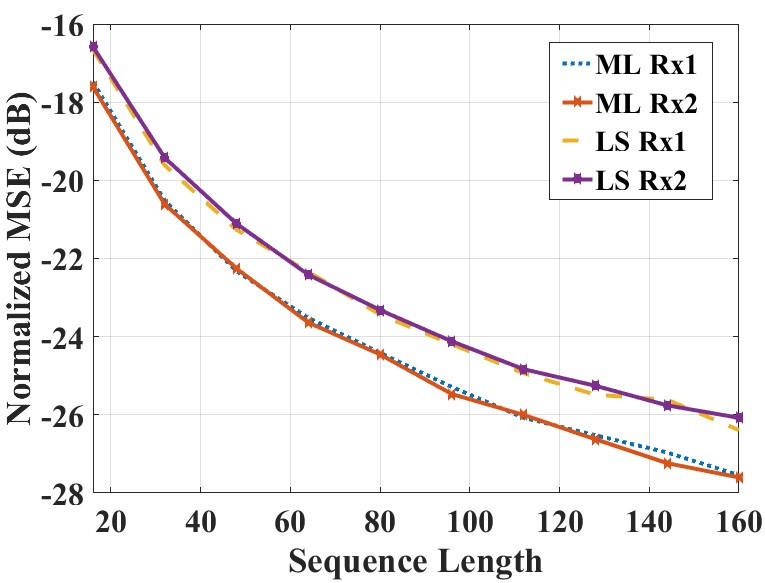}\\
  \caption{Normalized MSE in dB vs. training sequence (K) for ML and LS estimators with L = 3}\label{norm_MSE}
 
\end{figure}

In terms of mean square error (MSE), $\mathbbm{E} \, \big\{   \hat{\bar{\vect{c}}}_j -\bar{\vect{c}}_j ||^{\, 2}  \big\} $ in dB vs. the training sequence length (K) for the ML and LS (Fig. \ref{ML_LS}) estimators, respectively. The MSE decreases with increasing the training sequence length as we expected. We can notice that training sequences are designed such that both receivers have optimum performances for both estimators as they attain the corresponding CRB. Fig. \ref{norm_MSE}, shows the Normalized MSE which is defined by 
\begin{equation}
MSE_j^N=\frac{\mathbbm{E} \, \big\{  || \hat{\bar{\vect{c}}}_j -\bar{\vect{c}}_j ||^{\, 2}  \big\}}{|| \mathbbm{E}  \{ \bar{\vect{c}}_j   \} ||^{\, 2}}.
\end{equation}  

The value of the normalized MSE is much lower, around 38dB, than the MSE. As we can see in Fig. \ref{norm_MSE}, the performance of the ML estimator outperforms the LS estimator by approximately 1 dB. However, the LS estimator is preferred due to its simplicity respect to the ML estimator, because our bio-based receivers have limited computational capabilities. In applications where receivers send the data to the external computers, the ML estimator is preferred because it reaches to the CRB bound. 
 \section{Conclusions}
 In this paper, we have presented a training-based channel estimation for $M\times M$ D-MIMO system. We have developed ML and LS estimators which consider ISI of the pair transmitter-receiver and also inter-link diffusive interference. We have also derived the Cramér-Rao bound for each receiver. Training sequences have been designed such that minimize the CRB for all receivers simultaneously. The ML estimator outperforms the LS estimator and the corresponding MSE reaches the CRB. However, ML estimator needs to solve a system of non-linear equations and this makes it computationally not preferable. In the other hand, LS estimator is very simple to design at the price of a small performance degradation (1 dB for the example considered here) compared to the CRB.

\bibliographystyle{ieeetr}
\bibliography{Bibliography.bib}

\begin{thebibliography}{10}

\bibitem{nakano2005molecular}
T.~Nakano, T.~Suda, M.~Moore, R.~Egashira, A.~Enomoto, and K.~Arima,
  ``Molecular communication for nanomachines using intercellular calcium
  signaling,'' in {\em Nanotechnology, 2005. 5th IEEE Conference on},
  pp.~478--481, IEEE, 2005.

\bibitem{farsad2016comprehensive}
N.~Farsad, H.~B. Yilmaz, A.~Eckford, C.-B. Chae, and W.~Guo, ``A comprehensive
  survey of recent advancements in molecular communication,'' {\em IEEE
  Communications Surveys \& Tutorials}, vol.~18, no.~3, pp.~1887--1919, 2016.

\bibitem{akyildiz2015internet}
I.~Akyildiz, M.~Pierobon, S.~Balasubramaniam, and Y.~Koucheryavy, ``The
  internet of bio-nano things,'' {\em IEEE Communications Magazine}, vol.~53,
  no.~3, pp.~32--40, 2015.

\bibitem{nakano2014molecular}
T.~Nakano, T.~Suda, Y.~Okaie, M.~J. Moore, and A.~V. Vasilakos, ``Molecular
  communication among biological nanomachines: A layered architecture and
  research issues,'' {\em IEEE transactions on nanobioscience}, vol.~13, no.~3,
  pp.~169--197, 2014.

\bibitem{kadloor2012molecular}
S.~Kadloor, R.~S. Adve, and A.~W. Eckford, ``Molecular communication using
  brownian motion with drift,'' {\em IEEE Transactions on NanoBioscience},
  vol.~11, no.~2, pp.~89--99, 2012.

\bibitem{nakano2013molecular}
T.~Nakano, A.~W. Eckford, and T.~Haraguchi, {\em Molecular communication}.
\newblock Cambridge University Press, 2013.

\bibitem{kuran2011modulation}
M.~S. Kuran, H.~B. Yilmaz, T.~Tugcu, and I.~F. Akyildiz, ``Modulation
  techniques for communication via diffusion in nanonetworks,'' in {\em
  Communications (ICC), 2011 IEEE International Conference on}, pp.~1--5, IEEE,
  2011.

\bibitem{noel2014optimal}
A.~Noel, K.~C. Cheung, and R.~Schober, ``Optimal receiver design for diffusive
  molecular communication with flow and additive noise,'' {\em IEEE
  transactions on nanobioscience}, vol.~13, no.~3, pp.~350--362, 2014.

\bibitem{cobo2010bacteria}
L.~C. Cobo and I.~F. Akyildiz, ``Bacteria-based communication in
  nanonetworks,'' {\em Nano Communication Networks}, vol.~1, no.~4,
  pp.~244--256, 2010.

\bibitem{farsad2016capacity}
N.~Farsad, Y.~Murin, A.~Eckford, and A.~Goldsmith, ``On the capacity of
  diffusion-based molecular timing channels,'' in {\em Information Theory
  (ISIT), 2016 IEEE International Symposium on}, pp.~1023--1027, IEEE, 2016.

\bibitem{kim2014symbol}
N.-R. Kim, A.~W. Eckford, and C.-B. Chae, ``Symbol interval optimization for
  molecular communication with drift,'' {\em IEEE transactions on
  nanobioscience}, vol.~13, no.~3, pp.~223--229, 2014.

\bibitem{koo2016molecular}
B.-H. Koo, C.~Lee, H.~B. Yilmaz, N.~Farsad, A.~Eckford, and C.-B. Chae,
  ``Molecular mimo: From theory to prototype,'' {\em IEEE Journal on Selected
  Areas in Communications}, vol.~34, no.~3, pp.~600--614, 2016.

\bibitem{meng2012mimo}
L.-S. Meng, P.-C. Yeh, K.-C. Chen, and I.~F. Akyildiz, ``Mimo communications
  based on molecular diffusion,'' in {\em Global Communications Conference
  (GLOBECOM), 2012 IEEE}, pp.~5380--5385, IEEE, 2012.

\bibitem{jamali2016channel}
V.~Jamali, A.~Ahmadzadeh, C.~Jardin, H.~Sticht, and R.~Schober, ``Channel
  estimation for diffusive molecular communications,'' {\em IEEE Transactions
  on Communications}, vol.~64, no.~10, pp.~4238--4252, 2016.

\bibitem{arjmandi2013diffusion}
H.~Arjmandi, A.~Gohari, M.~N. Kenari, and F.~Bateni, ``Diffusion-based
  nanonetworking: A new modulation technique and performance analysis,'' {\em
  IEEE Communications Letters}, vol.~17, no.~4, pp.~645--648, 2013.

\bibitem{noel2014unifying}
A.~Noel, K.~C. Cheung, and R.~Schober, ``A unifying model for external noise
  sources and isi in diffusive molecular communication,'' {\em IEEE Journal on
  Selected Areas in Communications}, vol.~32, no.~12, pp.~2330--2343, 2014.

\end{thebibliography}
\end{document}